\documentclass[sigconf]{acmart}

\usepackage{multirow}
\usepackage{multicol}
\usepackage{subfigure}
\usepackage{relsize}

\usepackage{balance}

\copyrightyear{2026}
\acmYear{2026}
\setcopyright{cc}
\setcctype{by}
\acmConference[WSDM '26]{Proceedings of the Nineteenth ACM International Conference on Web Search and Data Mining}{February 22--26, 2026}{Boise, ID, USA}
\acmBooktitle{Proceedings of the Nineteenth ACM International Conference on Web Search and Data Mining (WSDM '26), February 22--26, 2026, Boise, ID, USA}
\acmPrice{}
\acmDOI{10.1145/3773966.3777970}
\acmISBN{979-8-4007-2292-9/2026/02}

\begin{document}

\title{Unlocking Scaling Law in Industrial Recommendation Systems with a Three-step Paradigm based Large User Model}



\author{Bencheng Yan}
\authornote{These authors contributed equally to this work and are co-first authors.}
\email{bencheng.ybc@alibaba-inc.com}
\affiliation{%
  \institution{Alibaba Group}
  \city{Beijing}
  \country{China}
}

\author{Shilei Liu}
\authornote{These authors contributed equally to this work and are co-first authors.}
\email{liushilei.lsl@alibaba-inc.com}
\affiliation{%
  \institution{Alibaba Group}
  \city{Beijing}
  \country{China}
}

\author{Zhiyuan Zeng}
\authornote{These authors contributed equally to this work and are co-first authors.}
\email{zengzhiyuan.zzy@alibaba-inc.com}
\affiliation{%
  \institution{Alibaba Group}
  \city{Beijing}
  \country{China}
}

\author{Zihao Wang}
\email{wzh454725@alibaba-inc.com}
\affiliation{%
  \institution{Alibaba Group}
  \city{Beijing}
  \country{China}
}

\author{Yizhen Zhang}
\email{zhangyizhen.zyz@alibaba-inc.com}
\affiliation{%
  \institution{Alibaba Group}
  \city{Beijing}
  \country{China}
}

\author{Yujin Yuan}
\email{yujin.yyj@alibaba-inc.com}
\affiliation{%
  \institution{Alibaba Group}
  \city{Beijing}
  \country{China}
}

\author{Langming Liu}
\email{liulangming.llm@alibaba-inc.com}
\affiliation{%
  \institution{Alibaba Group}
  \city{Beijing}
  \country{China}
}

\author{Jiaqi Liu}
\email{ljq414468@alibaba-inc.com}
\affiliation{%
  \institution{Alibaba Group}
  \city{Beijing}
  \country{China}
}

\author{Di Wang}
\email{zhemu.wd@alibaba-inc.com}
\affiliation{%
  \institution{Alibaba Group}
  \city{Beijing}
  \country{China}
}

\author{Wenbo Su}
\email{vincent.swb@alibaba-inc.com}
\affiliation{%
  \institution{Alibaba Group}
  \city{Beijing}
  \country{China}
}

\author{Wang Pengjie}
\email{pengjie.wpj@alibaba-inc.com}
\affiliation{%
  \institution{Alibaba Group}
  \city{Beijing}
  \country{China}
}

\author{Jian Xu}
\email{xiyu.xj@alibaba-inc.com}
\affiliation{%
  \institution{Alibaba Group}
  \city{Beijing}
  \country{China}
}

\author{Bo Zheng}
\authornote{Corresponding author.}
\email{bozheng@alibaba-inc.com}
\affiliation{%
  \institution{Alibaba Group}
  \city{Beijing}
  \country{China}
}

\renewcommand{\shortauthors}{Bencheng Yan et al.}

\begin{abstract}
Recent advancements in autoregressive Large Language Models (LLMs) have achieved remarkable progress, largely driven by their scalability—commonly formalized as the scaling law. 
Inspired by these successes, there has been growing interest in adapting LLMs to recommendation systems (RecSys) by reformulating recommendation tasks as generative sequence modeling problems.
However, existing End-to-End Generative Recommendation (E2E-GR) methods often sacrifice the practical advantages of traditional Deep Learning-based Recommendation Models (DLRMs)—including mature feature engineering, modular architectures, and production-grade optimization practices. 
This trade-off introduces critical challenges that hinder the effective application of scaling laws in industrial RecSys.
In this paper, we present Large User Model (LUM), a scalable and production-aware framework that bridges the gap between generative modeling and industrial recommendation requirements. 
LUM addresses these limitations through a principled three-step paradigm, designed to preserve the flexibility of autoregressive generation while maintaining compatibility with real-world deployment constraints.
Extensive experiments show that LUM outperforms state-of-the-art DLRMs and E2E-GR approaches across multiple benchmarks. Notably, LUM exhibits strong scalability: performance improves consistently as the model scales up to 7 billion parameters. Furthermore, LUM has been successfully deployed in a large-scale industrial application, where it delivered statistically significant gains in a live A/B test, demonstrating both its effectiveness and practical viability.
\end{abstract}



\maketitle

\begin{figure*}[t]
\centering
\includegraphics[width = .8\textwidth]{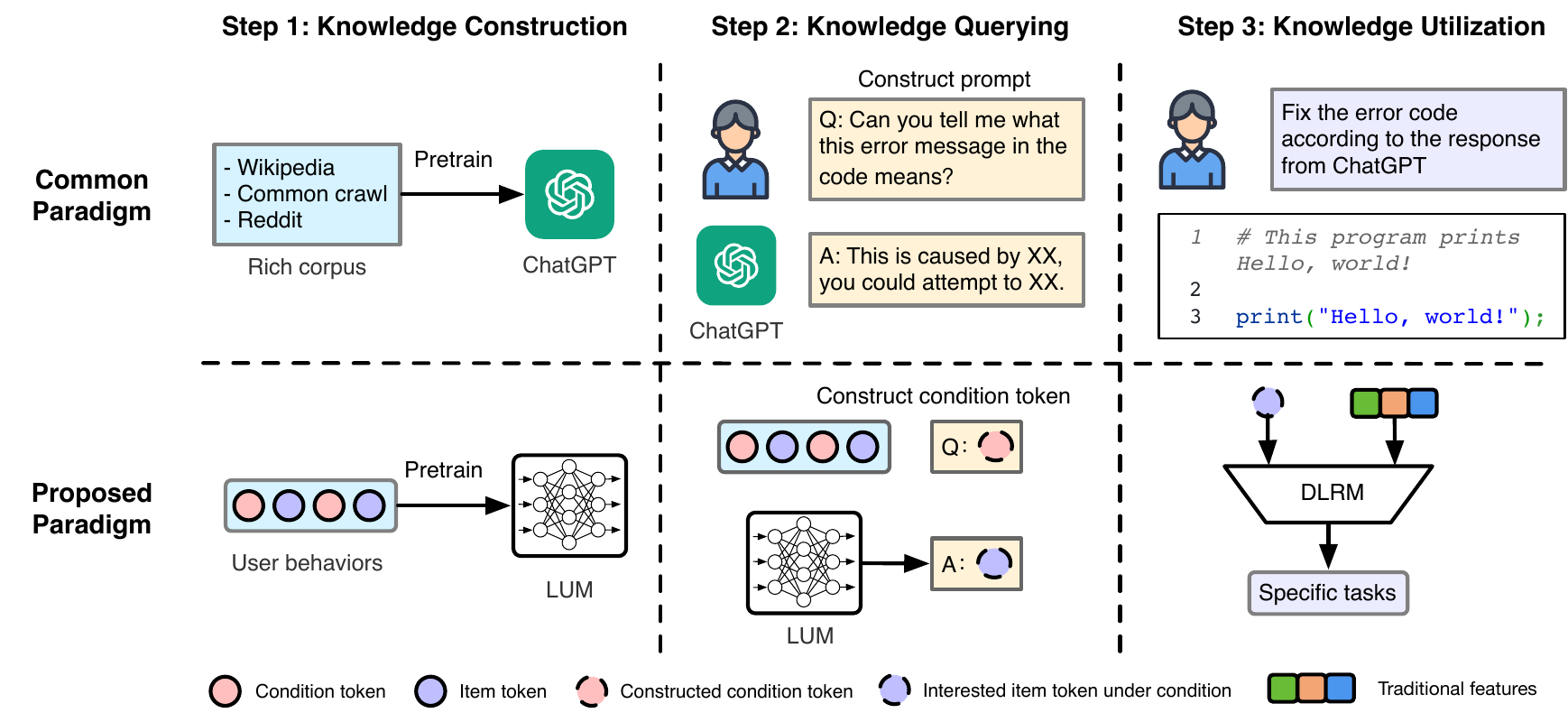}
\vspace{-1em}
\caption{Intuitive insight from the common paradigm in using LLM to the proposed \textbf{multi-step, generative-to-discriminative} paradigm.}
\vspace{-1em}
\label{figure:The comparison cases between LLM and LUM.}
\end{figure*}

\vspace{-1em}
\section{Introduction}
\label{sec:Introduction}
In recent years, autoregressive large language models (LLMs) have achieved significant breakthroughs, primarily due to their adherence to the principle of scalability, often referred to as the "scaling law".
This law posits that there is a predictable and continuous improvement in model performance as the computational resources allocated to the model are increased\cite{radford2018improving,radford2019language}.

In parallel, the field of recommendation systems (RecSys) has been actively investigating the potential for leveraging similar scalability principles.
Early research has highlighted that traditional deep learning-based recommendation models (DLRMs), widely used in industry, do not exhibit the same level of scalability observed in LLMs\cite{zhai2024actions}.
This discrepancy can be attributed to the fundamental differences between generative models and discriminative models. 
Specifically, generative models require a more substantial capacity to effectively capture the complex joint probability distribution 
$p(x,y)$ of the data, whereas discriminative models, such as traditional DLRMs, focus on modeling the simpler conditional probability 
$p(y|x)$. 
As a result, the benefits of increased computational resources are less pronounced in discriminative models \cite{ding2024inductive,liu2024multi,zhai2024actions}.



Consequently, a growing body of work has sought to adapt the generative modeling paradigm to recommendation systems by reformulating retrieval and ranking as generative tasks. 
This is typically accomplished by treating User Behavior Sequences (UBS) as a behavioral corpus and training a Transformer-based architecture in an end-to-end manner to perform next-item prediction \cite{zhai2024actions,geng2022recommendation}.
However, these End-to-End Generative Recommendation methods (E2E-GRs) often disregard the well-established advantages of traditional Deep Learning-based Recommendation Models (DLRMs)—including rich feature engineering, modular architectures, and production-proven training practices. This design choice introduces a critical trade-off, leading to a range of practical challenges that limit their effectiveness and scalability in industrial settings.


\noindent \textbf{(1) Inconsistency Between Generative Training and Discriminative Application:}
Despite their ability to capture complex sequential patterns and underlying data distributions, E2E-GRs often underperform on discriminative downstream tasks—such as click-through rate (CTR) prediction—where both calibration and ranking performance are critical \cite{DBLP:conf/kdd/YanQWBN22}.
This performance gap arises from a fundamental misalignment: generative models are optimized to maximize data likelihood, focusing on accurately modeling the process of user behavior generation, rather than directly optimizing for discriminative objectives like prediction accuracy or ranking quality.
Consequently, while E2E-GRs excel at distributional modeling, they may fail to meet the stringent requirements of industrial recommendation pipelines, which demand highly calibrated probabilities and fine-grained ranking sensitivity \cite{ng2001discriminative,bernardo2007generative,yogatama2017generative}.


\noindent \textbf{(2) Efficiency Challenges:}
The demand for high efficiency in continuous streaming training, coupled with stringent low-latency requirements for online real-time inference—extending beyond the throughput considerations discussed in \cite{zhai2024actions}—presents significant challenges for the direct deployment of E2E-GRs in industrial settings.
Even when deployment is feasible, these efficiency constraints fundamentally limit model scalability, particularly in terms of both model capacity and input sequence length (see Section~\ref{sec:Efficiency Evaluation}).


\noindent \textbf{(3) Lack of Flexibility:}
E2E-GRs exhibit a notable limitation in their ability to adapt to the dynamic and evolving nature of business requirements. Specifically, the incorporation of new types of behavioral data, such as refund behaviors or behaviors associated with new scenarios, presents a significant challenge. 
The inherent structure of E2E-GRs necessitates that any modification to the input schema, including the addition of new elements, triggers a requirement for retraining the entire model. 
This rigidity not only complicates the integration of new features but also imposes substantial constraints on the system's adaptability. 
The need to retrain the model from scratch upon any schema alteration is both time-consuming and resource-intensive, thereby diminishing the practicality of E2E-GRs in industrial settings where rapid and flexible responses to changing conditions are often critical.


\noindent \textbf{(4) Limited Compatibility.}
E2E-GRs are typically built on raw UBS using a Transformer architecture.
This design inherently limits their compatibility with established industrial practices—such as explicit feature engineering and parameter inheritance from deployed DLRMs—hindering seamless integration into production systems.
As a result, a significant performance gap often arises between the development phase of E2E-GRs and their real-world deployment (Section~\ref{sec:Effectiveness Evaluation}).
This gap is particularly evident in applications where the online model has been iteratively refined over years or even decades.
Such a discrepancy highlights the need for frameworks that better bridge the divide between theoretical advances and industrial practice.


In this paper, we rethink the critical question: \textbf{How can we effectively harness generative models to unlock scaling law in industrial settings?} This question compels us to revisit the common paradigm in using LLMs (Figure \ref{figure:The comparison cases between LLM and LUM.}).
Specifically, LLMs are initially trained in a generative manner on vast datasets, enabling them to assimilate a broad spectrum of knowledge.
End-users then interact with these models through diverse prompts, querying them across a wide array of topics. 
Ultimately, the responses generated by LLMs are utilized by end-users to make informed decisions across multiple domains.
This \textbf{multi-step, generative-to-discriminative} nature of LLMs provides a foundational framework for leveraging generative models in practical applications. 
Drawing upon this insight, we propose a three-step paradigm for training a Large User Model (LUM) tailored for industrial use. 
These steps are outlined as follows (see Figure \ref{figure:The comparison cases between LLM and LUM.}):
(1) \textbf{Step 1: Knowledge Construction.} 
A LUM is introduced, utilizing a transformer architecture and pre-trained through generative learning. 
This model captures user interests and the collaborative relationships among items, thereby characterizing a comprehensive knowledge base.
(2) \textbf{Step 2: Knowledge Querying.}
In this phase, LUM is queried with predefined questions pertaining to user-specific information, facilitating the extraction of relevant insights.
Intuitively, this process can be fundamentally conceptualized as a form of "prompt engineering" which is specifically designed to elicit an extensive knowledge.
(3) \textbf{Step 3: Knowledge Utilization.}
The outputs from LUM, obtained in Step 2, serve as supplementary features. 
These are integrated into traditional DLRMs to enhance their predictive accuracy and decision-making capabilities.

Generally speaking, we can benefit the three-step paradigm from 
(1) 
In the first step, the generative learning of LUM enables the exploration of scaling laws, which are critical for enhancing model performance.
(2) 
The decoupled designed of our paradigm eliminates the constraints associated with continue streaming training or serving. 
This separation facilitates the implementation of caching strategies for the LUM, thereby mitigating efficiency limitations.
(3) The third step ensures that DLRMs can meet the requirements for real-time learning, flexibility, and compatibility. 
This is achieved by integrating the previously learned LUM with DLRMs, thereby enhancing their adaptability to dynamic environments and ensuring seamless integration with existing systems.

The remaining challenge is to effectively transfer the learned data joint distribution $p(x,y)$ from the first step to downstream discriminative tasks, thereby addressing the aforementioned limitation. 
Ideally, the user knowledge encapsulated within LUM should exhibit a substantial correlation with the discriminative tasks at hand. 
To achieve this, we introduce a novel tokenization strategy for UBS, wherein each item is expanded into two distinct tokens: a condition token and an item token (Figure \ref{figure:An example of tokenization} and Section \ref{sec:Tokenization}).
Subsequently, we redefine the autoregressive learning process of UBS from \textit{"next-item prediction"}  to \textit{"next-condition-item prediction"}. 
This reformulation enables us to seamlessly \textbf{trigger} the relevant knowledge from the LUM into the discriminative tasks by specifying various conditions during the second step of the process. 
Finally, we have applied this approach to both offline datasets and online industrial applications, achieving significant improvements (Section \ref{sec:Performance on  Recommendation tasks} and \ref{sec:Online Results}).
Furthermore, the LUM demonstrates scaling properties similar to those observed in LLMs, allowing it to be successfully scaled up to 7 billion parameters while maintaining consistent performance enhancements. 
These findings underscore the robustness and adaptability of the LUM in diverse industrial application scenarios.

In summary, our contribution are summarized as follows:
(1) We introduce a pioneering three-step paradigm specifically tailored for industrial applications. 
(2) We propose a large user model (LUM) that incorporates a "next-condition-item prediction" task, to bridge the gap between generative pre-training and discriminative applications.
(3) Comprehensive empirical evaluations demonstrate that the proposed three-step paradigm based LUM significantly enhances the performance of downstream tasks in both public and industrial settings. 
Notably, our experiments reveal a clear scaling law governing the performance of LUM, which underscores the importance of scaling up the model size for optimal results. 
Additionally, we have successfully deploy it on our industrial applications and achieve significant gains in an A/B test.

\vspace{-1em}
\section{Related works}
\label{sec:Related works}
\noindent \textbf{Deep Learning based Recommendation Models.}
Traditional DLRMs typically use deep neural networks and fall into two categories:
(1) Retrieval-oriented models, such as two-tower architectures (e.g., EDB~\cite{huang2020embedding}) or sequential models (e.g., SASRec~\cite{kang2018self}, BERT4Rec~\cite{sun2019bert4rec}, GRU4Rec~\cite{hidasi2015session}), which capture user-item relevance;
(2) CTR prediction models, often based on Embedding + MLP structures~\cite{guo2017deepfm,lian2018xdeepfm,zhou2018deep,zhou2019deep,pi2020search,chang2023twin,yan2022apg}.
Despite their success, these models struggle to scale with compute and fail to harness advances in large foundation models.

\noindent \textbf{Generative Recommendation Models.}
To study scaling laws, prior work has adapted LLM-like autoregressive transformers for next-item prediction~\cite{zhai2024actions,geng2022recommendation}.
However, as noted in Section~\ref{sec:Introduction}, these methods rely on idealized assumptions and neglect the strengths of traditional DLRMs in features, architecture, and industrial practices—limiting their real-world applicability.
We address this with a three-step paradigm based on LUM.
While some recent efforts leverage LLMs’ open-world knowledge for content-based recommendations via end-to-end or multi-step training~\cite{bao2023tallrec,lin2024clickprompt,yu2024ra,chen2024hllm}, our focus remains on scalable modeling of collaborative signals, not content.

\section{Preliminary}
\subsection{Traditional DLRMs}
\label{sec:Traditional Deep Learning Methods in RecSys}
In the domain of RecSys, two primary tasks are identified: retrieval and ranking.
For the purpose of this discussion, we will focus on a search scenario, though it is important to note that the paradigm proposed herein is equally applicable to other industrial applications.
Given a user $u \in \mathcal{U}$, an item $i \in \mathcal{I}$, a search term $s \in \mathcal{S}$:

\noindent \textbf{Retrieval task:}
This task is aimed at identifying a subset of candidate items from the corpus that align with the user's $u$'s preferences, as influenced by the query $s$. 
A conventional approach to the retrieval task employs a two-tower architecture \cite{huang2020embedding}, comprising a user-query tower $\textit{UEnc}$ and an item tower $\textit{IEnc}$. 
These towers may consist of any suitable neural network structures, such as MLPs. 
The user-query tower encodes the user $u$ and the search term $s$ into a unified embedding $e^r_{us}=\textit{UEnc}(us)$, while the item tower generates an embedding for the item $i$ denoted as $e^r_i=\textit{IEnc}(i)$. 
Subsequently, contrastive learning techniques are utilized to refine the representations produced by these two towers.

\noindent \textbf{Ranking Task:} 
In contrast, the ranking task focuses on forecasting the likelihood of a user $u$ clicking on an item $i$ in response to a specific query $s$.
This is mathematically formalized as:
$\hat{y}=f(u,i,s)$ where $\hat{y}$ represents the predicted CTR and $f$ refers to the Embedding+MLP architecture.

\subsection{E2E-GRs via Next-item Prediction}


In the context of E2E-GRs, given a UBS for user $ u $, denoted as $ B_u = \{i_1, i_2, \dots, i_L\} $, where $ L $ represents the length of the sequence, the next-item prediction framework assumes that the probability of the next item $ i_k $ is conditionally dependent on the preceding items $ \{i_1, i_2, \dots, i_{k-1}\} $.  
Accordingly, the likelihood of the entire UBS $ B_u $ can be factorized as:  
$
p(i_1, i_2, \dots, i_L) = \prod_{l=1}^{L} p(i_l \mid i_1, i_2, \dots, i_{l-1}).
$
The objective of the autoregressive learning paradigm in E2E-GRs is to model and optimize the joint distribution $ p_\theta(i_1, i_2, \dots, i_L) $, a task commonly referred to as "next-item prediction"\footnote{
It is worth noting that some approaches extend this formulation by jointly modeling the types of user actions (e.g., click, purchase, view)~\cite{zhai2024actions}. However, the underlying modeling principle remains fundamentally similar.}.

\begin{figure}[t]
\centering
\includegraphics[width = .3\textwidth]{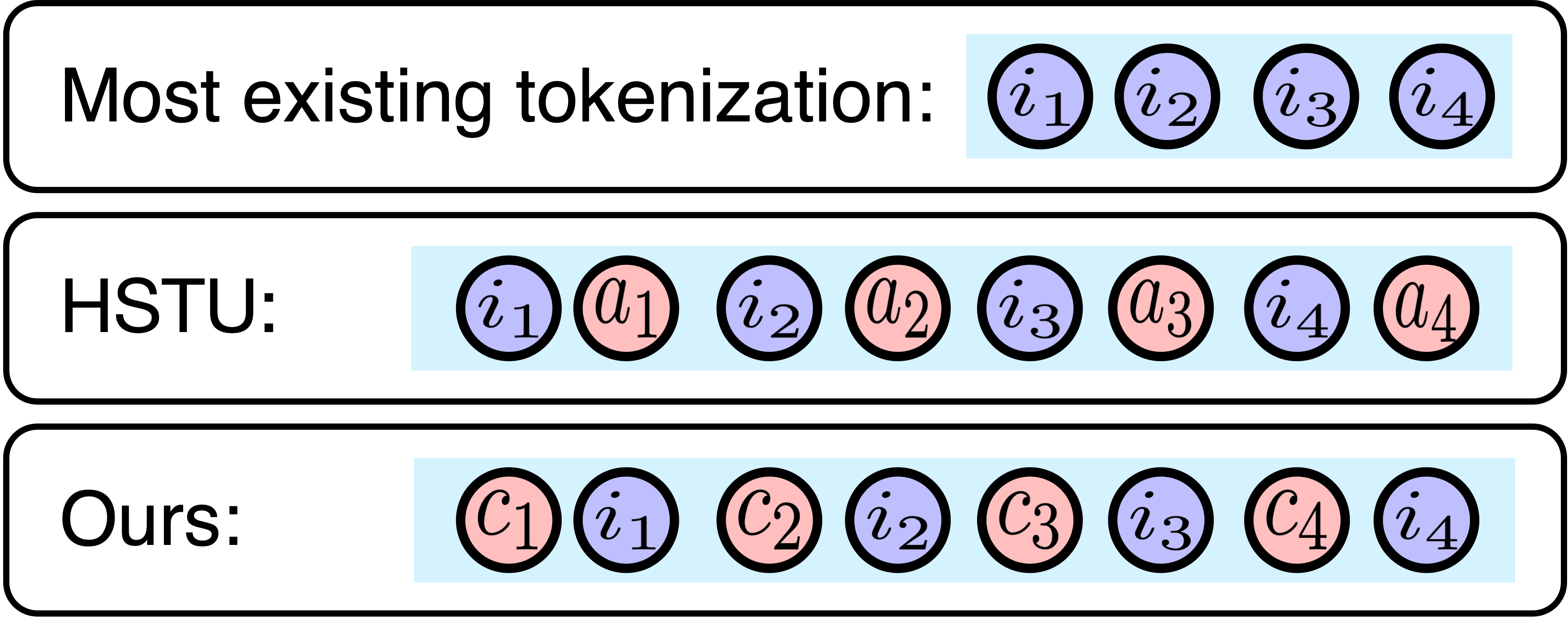}
\vspace{-1em}
\caption{The comparison of different tokenizations}
\vspace{-1em}
\label{figure:An example of tokenization}
\end{figure}

\begin{figure*}[t]
\centering
\includegraphics[width = .85\textwidth]{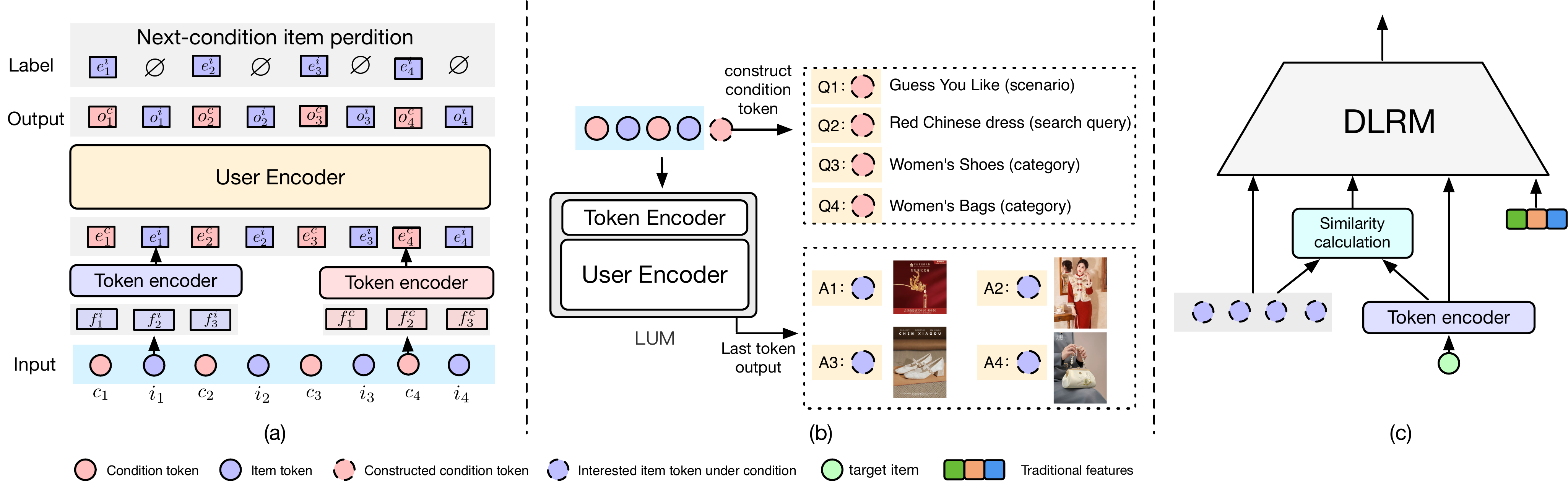}
\vspace{-1em}
\caption{
The three-step paradigm based large user model.
}
\vspace{-1em}
\label{figure:The architecture of LUM.}
\end{figure*}

\section{Method}
\label{sec:Method}
\subsection{Step 1: Knowledge Construction via Pre-training LUM}
\subsubsection{Tokenization}
\label{sec:Tokenization}

In this work, we propose a novel reformulation of autoregressive modeling in UBS, shifting from the conventional "next-item prediction" framework to a more expressive "next-condition-item prediction" paradigm.  
Specifically, each item $ i_k $ in the sequence $ B_u = \{i_1, \dots, i_L\} $ is decomposed into two distinct tokens: a condition token $ c_k $ and an item token $ i_k $. As a result, the UBS is transformed into an alternating sequence:  
$
\{c_1, i_1, c_2, i_2, \dots, c_L, i_L\},
$ where $ c_k $ encodes contextual signals associated with the subsequent item $ i_k $ (see Figure~\ref{figure:An example of tokenization}).
To illustrate, consider a user who interacts with items across multiple application scenarios—such as search and recommendation. In this case, $ c_k $ can be instantiated as a scenario token (e.g., "search" or "homepage recommendation"), explicitly modeling the context in which $ i_k $ was engaged. 
This design addresses a key limitation of standard autoregressive models: without explicit conditioning, generated items reflect a general interest distribution that may be misaligned with downstream tasks. 
For instance, considering a search scenario, after a user issues the query "dresses", a model lacking condition control may generate recommendations unrelated to fashion or intent. 
In contrast, our framework enables explicit control over generation behavior: during inference, the condition token acts as a trigger that steers the model toward task-specific outputs by setting appropriate context. 
This mechanism is conceptually analogous to prompt engineering in large language models, where different prompts elicit distinct knowledge patterns.


Notably, while HSTU~\cite{zhai2024actions} also uses auxiliary tokens (e.g., action types), its modeling differs fundamentally: each interaction is a tuple $\langle i_k, a_k \rangle$, where action $a_k$ (e.g., click) is tied to the \emph{past} item $i_k$ (Figure~\ref{figure:An example of tokenization}), not used to condition future predictions.  
This limits its ability to dynamically shape the next-item distribution based on prospective context.  
In contrast, our formulation supports context-aware generation and task-controllable inference, significantly enhancing E2E-GR flexibility in practice.

\vspace{-1em}
\subsubsection{Architecture}
\label{sec:Architecture}
The overarching framework of LUM is depicted in Figure \ref{figure:The architecture of LUM.} (a), presenting a hierarchical structure that encompasses both a Token Encoder and a User Encoder.

\noindent \textbf{Token Encoder.}
The input tokens in LUM are inherently heterogeneous, falling into two primary types: condition tokens ($c$) and item tokens ($i$). 
Each token type is associated with a set of attribute features—for instance, item tokens may include ID-based (e.g., item ID, brand ID), statistical (e.g., click-through rate), and content-derived (e.g., title embedding) features, while condition tokens may incorporate contextual metadata such as scenario ID or query text.  
To unify these heterogeneous inputs, we design a token encoder that maps each token into a shared embedding space. The encoding process first concatenates the feature representations of a token, followed by a projection layer that transforms the concatenated vector into a compact, dense embedding. 
Formally, the embedding $ e^t $ of token type $ t \in \{i, c\} $ is computed as:
$e^t = \textit{proj}^t(\textit{concat}(f^t_1;f^t_2;f^t_3;\dots)); t \in \{i,c\}$
where $f^t_k$ denotes the $k$-th feature representation of token $t$, and $ \textit{proj}^t $ is a learnable transformation. In this work, $ \textit{proj}^t $ is implemented as a linear layer unless otherwise specified.

\noindent \textbf{User Encoder:}
The user encoder is structured to capture user preferences and the collaborative information among items.
Specifically, the sequence of input tokens $\{c_1,i_1,c_2,i_2,\dots,c_L,i_L\}$ is represented as $\{e^c_1,e^i_1,e^c_2,e^i_2,\dots,e^c_L,e^i_L\}$ through the token encoder.
Subsequently, as illustrated in Figure \ref{figure:The architecture of LUM.} (a), the user encoder utilizes a conventional autoregressive transformer architecture to process these embeddings. 
The final output of the user encoder is denoted as $o^c_k$, encapsulating the integrated information from the input sequence.

\subsubsection{Next-condition-item Prediction}
Unlike next-item prediction, which focuses on predicting the subsequent item directly, next-condition-item prediction is concerned with predicting the next item given a specific condition. 
This approach necessitates the application of an autoregressive loss solely on the output of the condition token to infer the next item. Consequently, the autoregressive likelihood for next-condition-item prediction can be formulated as follows: $p(c_1,i_1,c_2,i_2,\dots,c_L,i_L) = \prod_{l=1}^{L} p(i_l|c_1,i_1,c_2,i_2,\dots,i_{l-1},c_l)$
Furthermore, to enhance the optimization of $p_{\theta}(c_1,i_1,c_2,i_2,\dots,c_L,i_L)$ in practical industrial applications, we employ the InfoNCE loss function \cite{oord2018representation} and introduce a packing strategy. 

\noindent \textbf{InfoNCE Loss.} 
In industrial applications, the vocabulary size of items can scale to billions, rendering the direct computation of generative probabilities over the entire set of items impractical. 
To address this challenge, we employ the InfoNCE  loss for predicting the next conditional items. 
The InfoNCE loss can be mathematically formulated as follows: $Loss = -\sum_{l=1}^{L} \log \left( \frac{\exp(sim(o^c_{l}, e^i_{l}) ))}{\exp(sim(o^c_{l}, e^i_{l})+\sum_{k=1}^{K}\exp(sim(o^c_{l}, e^i_{k}) ))} \right)$
where $sim$ is the similarity function.
For each item $l$, the other items within the same batch serve as negative samples and $K$ is the number of negative items.
$e^i_{k}$ is the embedding of $k$-th negative items.

\noindent \textbf{Packing.}
In practical applications, the lengths of UBSs exhibit significant variability among users. 
Indeed, the majority of UBS lengths are substantially shorter than the predefined maximum length. 
Processing each UBS individually in such scenarios is computationally inefficient. 
Drawing inspiration from the packing strategies employed in the GPT series \cite{radford2018improving, radford2019language, zhao2024analysing}, we adopt a similar approach by grouping multiple UBSs into a single sequence, thereby maximizing the utilization of the available sequence length.

\vspace{-1em}
\subsection{Step 2: Knowledge Querying with Given Conditions}
\label{sec:Step 2: Knowledge Querying with Given Conditions}

In Step 1, we construct the joint probability distribution over the tokenized sequence:
$ p(c_1, i_1, c_2, i_2, \dots, c_L, i_L),$ as introduced in Section~\ref{sec:Tokenization}. 
The subsequent step involves querying task-relevant knowledge from this distribution through conditional generation. 
Specifically, for each downstream task request, we first construct a corresponding condition token $ c_q $ based on the query context. 
For example, if the request corresponds to a user searching for items related to the query “dresses” in a search scenario, we can set $ c_q $ to include the associated scenario ID and query text. 
Given $ c_q $, the conditional probability
$ p(i_q \mid c_1, i_1, c_2, i_2, \dots, c_L, i_L, c_q)$ is computed to estimate the likelihood of user interest in item $ i_q $ (see Figure~\ref{figure:The architecture of LUM.}(b)).  

Notably, this mechanism—using condition tokens as triggers to elicit specific knowledge—establishes a bridge between generative sequence modeling and discriminative downstream tasks, thereby improving adaptability and effectiveness in real-world applications.  
 We illustrate this capability through the following examples:

\noindent $\bullet$ \textit{Example 1.} When $ c_q $ encodes a scenario identifier (e.g., “search” or “homepage”), the model infers user interests under different application contexts.

\noindent $\bullet$ \textit{Example 2.} If $ c_q $ represents a search query, the model generates recommendations conditioned on fine-grained user intent.

\noindent $\bullet$ \textit{Example 3.} When $ c_q $ denotes a category label, the model captures category-specific preferences.
 
Moreover, the framework supports multi-condition input by incorporating multiple feature fields into the condition token (i.e., extending the set $\{f^c_1;f^c_2;f^c_3;\dots\}$), enabling joint conditioning on scenario, intent, and metadata. Empirical results show that integrating diverse conditions significantly improves performance (see Section~\ref{sec:Effectiveness Evaluation}).  
Conceptually, this process mirrors \textbf{prompt engineering} in large language models, where carefully designed prompts activate relevant knowledge patterns—here, condition tokens serve as structured prompts to steer the generative model toward task-specific outputs.

It is noteworthy that, in Step 2, the condition token may encounter cold-start issues when dealing with new scenarios or new categories, which cannot be directly triggered. As for new queries, since we obtain embeddings from tokenized query text as features for the condition token, the textual features inherently possess a certain level of generalization to alleviate the cold-start problem. 
Considering the relatively low frequency of new scenarios and categories in industrial applications, along with the flexibility to adjust the contextual content of the condition token, the cold-start issue is not the focus of this study and will be left for future work.

\begin{figure}[t]
\centering
\includegraphics[width = .35\textwidth]{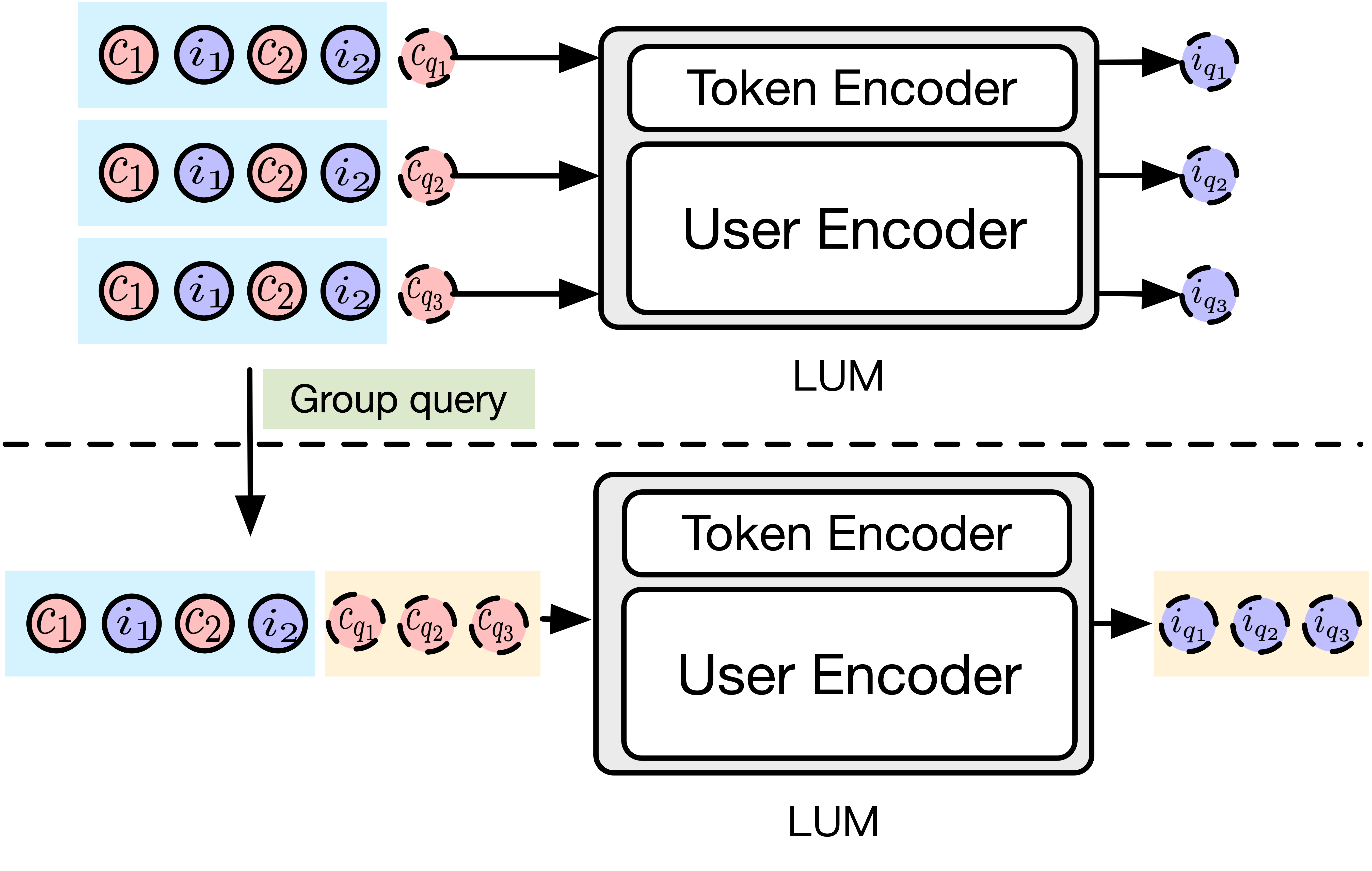}
\vspace{-1em}
\caption{An example of group query.}
\vspace{-2em}
\label{figure:An example of group query.}
\end{figure}

\noindent \textbf{Group query for efficiency.}
Moreover, given that a single user may response multiple queries, each pertaining to the same UBS, processing these queries in isolation can lead to significant inefficiencies. 
This issue is exacerbated in practical scenarios where the number of users can easily scale to billions, leading to an extensive number of <user, query> pairs requiring inference. 
To mitigate this challenge, we introduce a novel group query strategy aimed at enhancing computational efficiency.
As illustrated in Figure \ref{figure:An example of group query.}, all queries are concatenated into a single sequence, represented as $p(i_{q_1},i_{q_2},\dots|c_1,i_1,c_2,i_2,\dots,c_L,i_L,c_{q_1},c_{q_2},\dots)$.
To ensure that the inference process remains coherent and accurate, we apply a masking mechanism to prevent attentional interactions between different query condition $c_{q_j}$.
This approach allows the common prefix $\{c_1,i_1,c_2,i_2,\dots,c_L,i_L\}$ of various queries to be computed only once, while simultaneously querying the items $i_{q_j}$ under different conditions.
Empirical evaluations demonstrate that, with the implementation of the group query strategy, the inference process can be significantly accelerated by 78\%(see Section \ref{sec:Effectiveness Evaluation}).

\subsection{Step 3: Knowledge Utilization in DLRMs}
\label{sec:Step 3: Utilize knowledge in DLRMs}
After Step 2, we acquire a set of $N$ next-condition items $\{i_{q_1},i_{q_2},..,i_{q_N}\}$, with their corresponding outputs represented as $o^i_{q_n}$.
Additionally, each item $i \in \mathcal{I}$ is encoded through the token encoder (Section\ref{sec:Architecture}), yielding an embedding denoted as $e^i_i$.

To enhance existing DLRMs, we propose two strategies (see Figure \ref{figure:The architecture of LUM.} (c)):
\noindent (1) Direct Feature Incorporation: We integrate the outputs $o^i_{q_n}$ and the embedding $e^i_i$ of target item $i$ as fixed additional features into the DLRMs.  
This approach leverages the rich representations derived from the next-condition items directly.
\noindent (2) Interest Matching via Similarity Measurement: We assess the alignment between the target item $i$ and user interests by computing the similarity $sim(o^i_{q_n},e^i_i)$.
This similarity score quantifies how well the target item matches the context provided by the next-condition items.

Formally, for retrieval tasks, the two-tower model can be reformulated as: $e^r_{us}=\textit{UEnc}(us,\{o^i_{q_1},o^i_{q_2},..,o^i_{q_N}\})$ and $e^r_i=\textit{IEnc}(i,e^i_i)$.
For ranking tasks, the ranking model can be rewritten as: $\hat{y}=f(u,i,s,\{o^i_{q_n},sim(o^i_{q_n},e^i_i)|n=1,..,N\},e^i_i\}$.
This framework effectively integrates contextual information and item embeddings to improve both retrieval and ranking performance in recommendation systems.

\subsection{Industrial Deployment}
\label{sec:Industrial Deployment}
\begin{figure}[t]
\centering
\includegraphics[width = .4\textwidth]{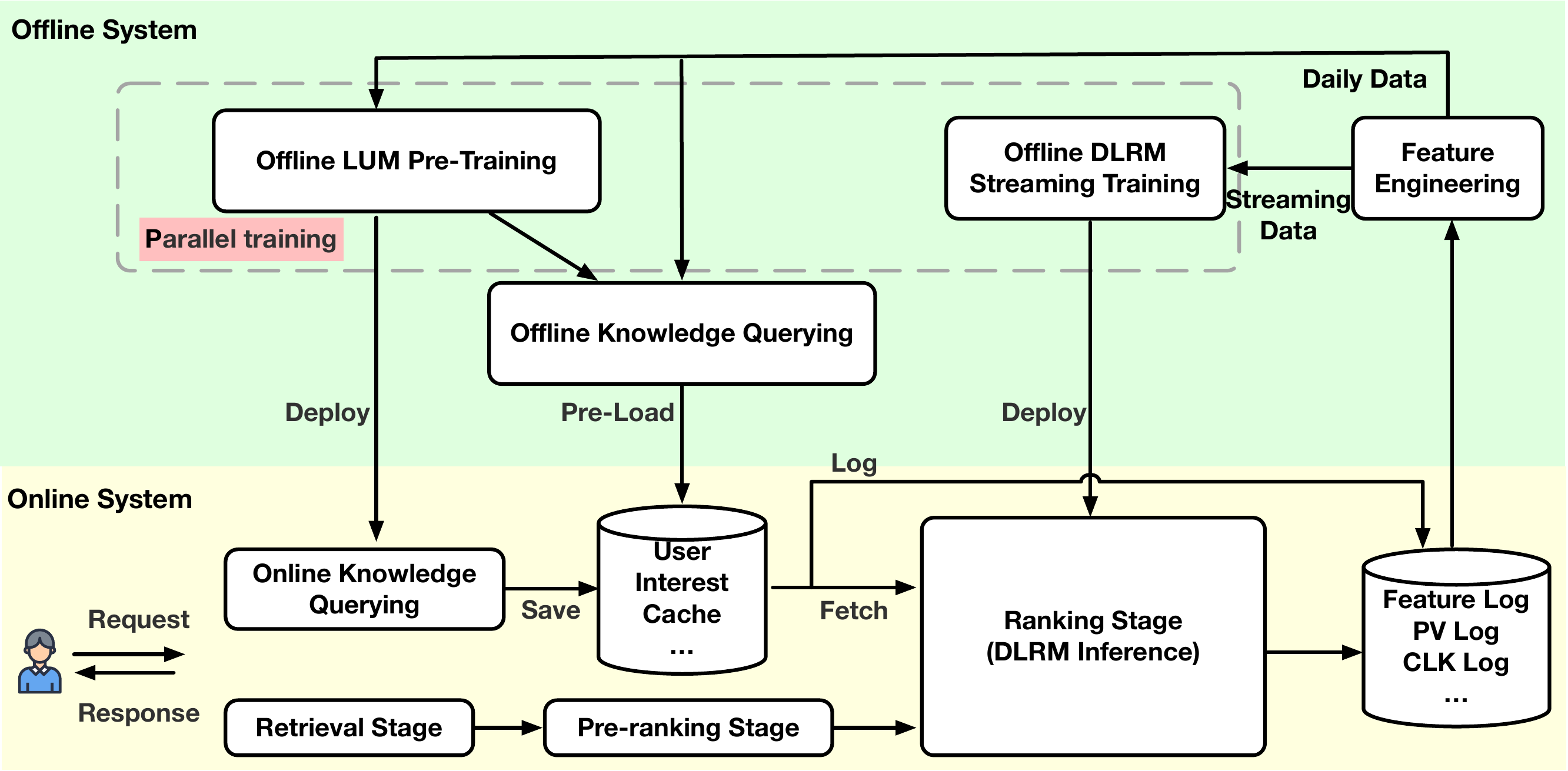}
\vspace{-1em}
\caption{Online deployment.}
\vspace{-2em}
\label{figure:online deployment.}
\end{figure}

In this section, we describe the deployment of LUM in Taobao’s sponsored search system (Figure~\ref{figure:online deployment.}), where LUM is primarily integrated into the ranking stage.
The core idea is based on the three-step paradigm: by decoupling the steps, we enable maximal parallelization across training and inference. \textbf{During online serving}, Step 2 (knowledge querying) is executed in real time during the early phase of user request processing—in parallel with retrieval and pre-ranking stage. This design decouples LUM inference from the strict latency constraints of the final ranking stage, allowing sufficient time for computation.
We build a centralized cache system that serves two purposes:
(1) it stores real-time inference results from knowledge querying;
(2) it pre-loads offline-inferred user interests as fallbacks to ensure coverage under tight RT (response time) constraints.
Then, the system retrieves the generated interests from the cache and feeds them to the DLRM in the ranking stage, while logging the results for offline training consumption.
\textbf{During offline training time}, we adopt a synchronous training pipeline that aligns LUM pre-training with DLRM streaming training.
For DLRM training, we directly consume logged user interests produced by online LUM inference—bypassing Step 1 and 2 entirely. This significantly improves training efficiency, a critical requirement for industrial-scale streaming systems.
For LUM, we train it in parallel with DLRM. Once updated, the new LUM model is deployed online, and its inference outputs are logged to influence subsequent DLRM training—forming a closed-loop, co-evolutionary learning framework.

\subsection{Discussion}
\label{sec:Discussion}
The proposed three-step paradigm based on LUM addresses the four limitations mentioned in Section \ref{sec:Introduction} as follows:


\noindent $\bullet$ \textbf{Addressing Limitation 1:}
To address the first limitation, we design a generative-to-discriminative learning pipeline.
In Step 1, the joint distribution $p(c_1,i_1,c_2,i_2,\dots,c_L,i_L)$ is learned via generative modeling over user behavior sequences.
In Step 2, a \textit{next-condition-item prediction} objective is introduced, enabling the model to activate relevant knowledge for downstream recommendation tasks.
Finally, in Step 3, discriminative learning is performed to align with the requirements of discriminative CTR applications.


\noindent $\bullet$ \textbf{Addressing Limitation 2:}  
Efficiency demands arise from the need for downstream models to meet strict efficiency constraints.  
For E2E-GR, this implies tight latency and throughput requirements on the model itself.  
In contrast, our three-step paradigm naturally decouples pre-training and knowledge querying (Steps 1–2) from downstream deployment (Step 3).  
Thus, only Step 3 must satisfy real-time inference constraints.  
As illustrated in Figure~\ref{figure:online deployment.}, during offline training, LUM’s pre-training and knowledge querying run in parallel with DLRM’s streaming training—where DLRM consumes the querying outputs from logs—introducing no additional overhead to DLRM’s training pipeline.  
During online serving, knowledge is pre-computed and cached, leaving DLRM's computation unaffected in ranking stage.

\noindent $\bullet$ \textbf{Addressing Limitation 3:}  
Like efficiency challenge, flexibility requirements are primarily imposed by downstream applications.  
For E2E-GR, both efficiency and flexibility must be jointly satisfied within a single model.  
In our framework, however, downstream flexibility needs are naturally handled by the DLRM in Step 3, highlighting the advantage of our decoupled design.  
Moreover, flexibility can also be supported during LUM pre-training: dynamic requirements (e.g., new user behaviors or business rules) can be unified as conditioning signals, consistent with our design of \textit{condition tokens}.  
That is, any evolving requirement can be consistently represented as a tuple $\langle\text{condition token}, \text{item token}\rangle$, enabling continuous adaptation of LUM through ongoing pre-training.  
Notably, unlike E2E-GR, LUM’s pre-training is not subject to strict efficiency constraints, allowing flexible integration of new conditions.

\noindent $\bullet$ \textbf{Addressing Limitations 4:}
In our framework, the DLRM in Step 3 serves as the online ranking model.
This design enables direct reuse of pre-existing industrial knowledge, including engineered features and learned representations.
Moreover, the system can seamlessly incorporate ongoing advancements in DLRM architectures and training techniques, ensuring long-term maintainability and performance evolution.

\section{Experiments}




\subsection{Experimental setting}

\begin{table}[t]
\center
\vspace{-1em}
\caption{The statistic of datasets.}
\vspace{-1em}
\tiny
\resizebox{\columnwidth}{!}{
\begin{tabular}{lccc}
\toprule
             & \textbf{\#Interaction}    & \textbf{\#User}   & \textbf{\#Item}   \\\midrule
Amazon Books & 914,014& 694,897& 686,623\\ 
MovieLen 1M (ML-1M) & 9,810,868&6,040&  3,883\\ 
MovieLen 20M (ML-20M) &  17,253,665 &138,493& 27,278 \\ 
Industrial Dataset & 4 billion & 0.1 billion & 0.1 billion  \\ 
\bottomrule
\end{tabular}
}
\label{table:The statistic of datasets}
\end{table}

\noindent \textbf{Datasets.}
In this study, we utilize three public datasets and one industrial dataset to evaluate the performance of our proposed method, LUM.
The public datasets include two benchmark datasets: MovieLens, which comprises two subsets (1M and 20M), and Amazon Books \cite{zhou2019deep,zhai2024actions}. 
The industrial dataset is sourced from the Taobao e-commerce platform.
The statistical information for these datasets is summarized Table \ref{table:The statistic of datasets}.

\noindent \textbf{Baselines.}
To comprehensively evaluate the performance of LUM, we compare it against a variety of state-of-the-art models. 
For the traditional retrieval model, we use the two-tower architecture-based EDB \cite{huang2020embedding}. 
For traditional ranking models, we consider DIN \cite{zhou2018deep}, DIEN \cite{zhou2019deep}, SIM \cite{pi2020search}, and TWIN \cite{chang2023twin}. 
Additionally, we compare LUM with E2E-GRs, specifically HSTU \cite{zhai2024actions}. 
We also include the traditional sequential recommendation model SASRec \cite{kang2018self}, which adopts a transformer architecture to model UBS, as a baseline.


\noindent \textbf{Training Configuration.}
To ensure a fair and rigorous comparison, we adopt a standardized training setup across all models.
By default, transformer-based models—including LUM, HSTU, and SASRec—are configured to have comparable model sizes. Specifically, on public benchmark datasets, the number of parameters is set to approximately 30 million; on the industrial dataset, it is scaled to 0.3 billion to reflect real-world modeling demands.
For other DLRM variants, model configurations follow the recommendations from their original papers. Notably, we try to increase the model size of these baselines but observe no further performance gains—consistent with findings reported in prior work~\cite{zhou2018deep}.
By default, all models are trained from scratch using an identical feature set to ensure a controlled evaluation.
The sequence length is set to 256 for public datasets and 4096 for the industrial dataset, reflecting differences in user behavior span and system requirements. 
The learning rate is set to 1e-3and 1e-4 for public and industrial datasets, respectively, and the batch size is fixed at 2048 for all experiments. 
Training is conducted using 32 GPUs for public datasets and 128 GPUs for the industrial dataset.
For LUM, the DLRM backbone in Step 3 is instantiated as SIM for ranking tasks and EDB for retrieval tasks on public datasets. 
On the industrial dataset, the backbone is set to the online model currently deployed in production, ensuring alignment with real-world serving infrastructure.

\begin{table}[]
\centering
\vspace{-1em}
\caption{Performance on Public Datasets}
\vspace{-1em}
\label{table:Ranking Performance on Public Datasets}
\tiny
\resizebox{0.9\columnwidth}{!}{
\begin{tabular}{lccc}
\toprule
\textbf{Model}     & \textbf{ML-1M} & \textbf{ML-20M} & \textbf{Amazon Books} \\ \midrule
SASRec    & 0.7295      & 0.7166       & 0.6699      \\  
HSTU      & 0.7533      & 0.7463      & 0.6712     \\
DIN       & 0.7455      & 0.7299       & 0.6139      \\
DIEN      & 0.7527      & 0.7319       & 0.6130      \\
SIM      & 0.7579      & 0.7341       & 0.6551      \\
TWIN      & 0.7539     & 0.7331       & 0.6538     \\ \midrule
LUM       & \textbf{0.7615}      & \textbf{0.7483}       &   \textbf{0.6727}           \\ \bottomrule
\end{tabular}
}
\vspace{-2em}
\end{table}

\subsection{Performance on  Recommendation tasks}
\label{sec:Performance on  Recommendation tasks}
\noindent \textbf{Performance on public datasets}
We first conduct experiments on public datasets to evaluate the performance of LUM. 
The performance metric reported is AUC\footnote{Note 0.001 absolute AUC gain is regarded as significant for the ranking task\cite{zhou2018deep}}. 
The results are summarized in Table \ref{table:Ranking Performance on Public Datasets}.
From the results, we observe that LUM achieves significant improvements across all datasets. 
This indicates that the three-step paradigm based LUM effectively captures a wide range of user interests and enhances the predictive accuracy of DLRMs.

\begin{table}[]
\centering
\caption{Overall performance in industrial settings.
Imp. denotes the improvements relative to the best baseline.}
\label{table:Overall Performance in Industrial Setting}
\vspace{-1em}
\tiny
\resizebox{0.9\columnwidth}{!}{
\begin{tabular}{lccc}
\toprule
\multirow{2}{*}{\textbf{Model}} & \textbf{Ranking} & \multicolumn{2}{c}{\textbf{Retrieval}} \\ 
                  & AUC     & R@10          & R@50         \\ \cmidrule{1-4}
SASRec            & 0.7322        &   0.2560        & 0.4740\\
DIN               & 0.7336       &     -            &    -          \\
HSTU              & 0.7334       &    0.2594        &   0.4781          \\
Online Model      & 0.7338       &    0.2482        &    0.4651          \\ \midrule
LUM               & \textbf{0.7514}      &     \textbf{0.2727}        &  \textbf{0.4915}            \\ 
Imp.               & \textbf{+0.0176}& \textbf{+0.0133}& \textbf{+0.0134}\\ \bottomrule
\end{tabular}
}
\end{table}

\noindent \textbf{Performance in industrial setting}
In this section, we compare the performance of LUM against both DLRMs and E2E-GRs in an industrial setting. 
For the baseline, we use the online model in our applications, which follows an Embedding+MLP architecture for ranking and a two-tower architecture for retrieval. 
We report AUC for ranking and Recall@K (R@K) for retrieval.
For LUM, the backbone DLRM in Step 3 is also set to the online model. 
Additionally, we include traditional state-of-the-art DLRMs (SASRec \cite{kang2018self} and DIN \cite{zhou2018deep}) and an E2E-GR (HSTU \cite{zhai2024actions}) as baselines for comparison. 
To ensure a fair comparison, all models use the same set of features. 
We set the maximum sequence length to 4096 for all models and train them from scratch.
The results are summarized in Table \ref{table:Overall Performance in Industrial Setting}. 
LUM achieves significant improvements over the best baseline, with a +0.0176 increase in AUC, a +0.0133 increase in R@10, and a +0.0134 increase in R@50. 
The substantial improvements can be attributed primarily to the generative-to-discriminative design of our proposed paradigm. 




\begin{table}[]
\centering
\caption{The performance of LUM on various DLRMs.
"Ave" denotes the average results across all cases, and "Imp." indicates the improvement of "Base+LUM" compared to "Base".
}
\vspace{-1em}
\label{table:The performance of LUM on various DLRMs.}
\resizebox{\columnwidth}{!}{
\begin{tabular}{llccccc}
\toprule
\multicolumn{2}{l}{}                            & DIN     & DIEN    & SIM     & TWIN   & Ave    \\ \midrule
\multirow{3}{*}{\textbf{ML-1M}}      & Base     & 0.7455  & 0.7527  &0.7579   &0.7539  & 0.7525 \\
                                     & Base+LUM & 0.7472  & 0.7604  & 0.7615  & 0.7675 & 0.7592 \\
                                     & Imp.     & \textbf{+0.0016}  & \textbf{+0.0078}  &  \textbf{+0.0036} & \textbf{+0.0136} & \textbf{+0.0067} \\\midrule
\multirow{3}{*}{\textbf{ML-20M}}     & Base     & 0.7299  & 0.7319  & 0.7341  & 0.7331 &  0.7323 \\
                                     & Base+LUM & 0.7413  &0.7361   & 0.7483 & 0.7422 & 0.7420 \\
                                     & Imp.     & \textbf{+0.0114}  &\textbf{+0.0042}   & \textbf{+0.0142} & \textbf{+0.0090}  &\textbf{+0.0097} \\ \midrule
\multirow{3}{*}{\begin{tabular}[c]{@{}l@{}}\textbf{Amazon} \\ \textbf{Book}\end{tabular}} 
                                     & Base     & 0.6139 &0.6130 & 0.6551 & 0.6538&  0.6340 \\
                                      & Base+LUM & 0.6261 & 0.6194 & 0.6727&  0.6591 & 0.6443      \\
                                      & Imp.    &  \textbf{+0.0122} & \textbf{+0.0063} & \textbf{+0.0176} & \textbf{+0.0053} & \textbf{+0.0103}       \\ 
                             \bottomrule
\end{tabular}
}
\end{table}

\begin{figure*}[t]
 \centering
     \subfigure[Model size (B) vs. Train time cost (hour)]{
    \includegraphics[width= .3\textwidth]{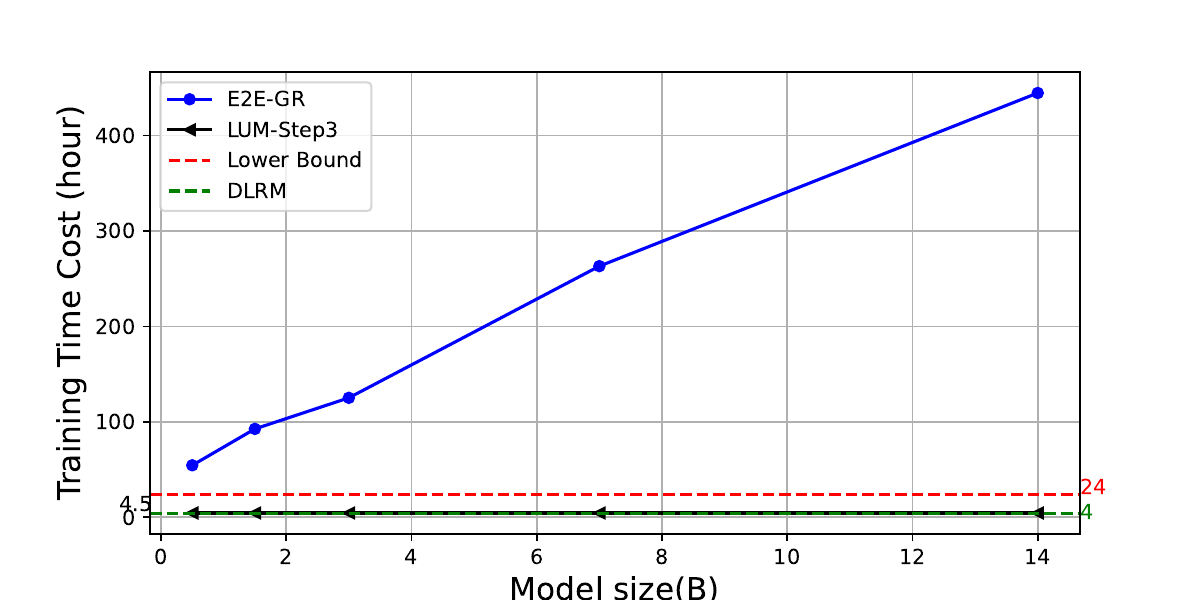}
    \label{fig:subfigure2}
  }
      \subfigure[Model size (B) vs. Latency (ms)]{
    \includegraphics[ width= .3\textwidth]{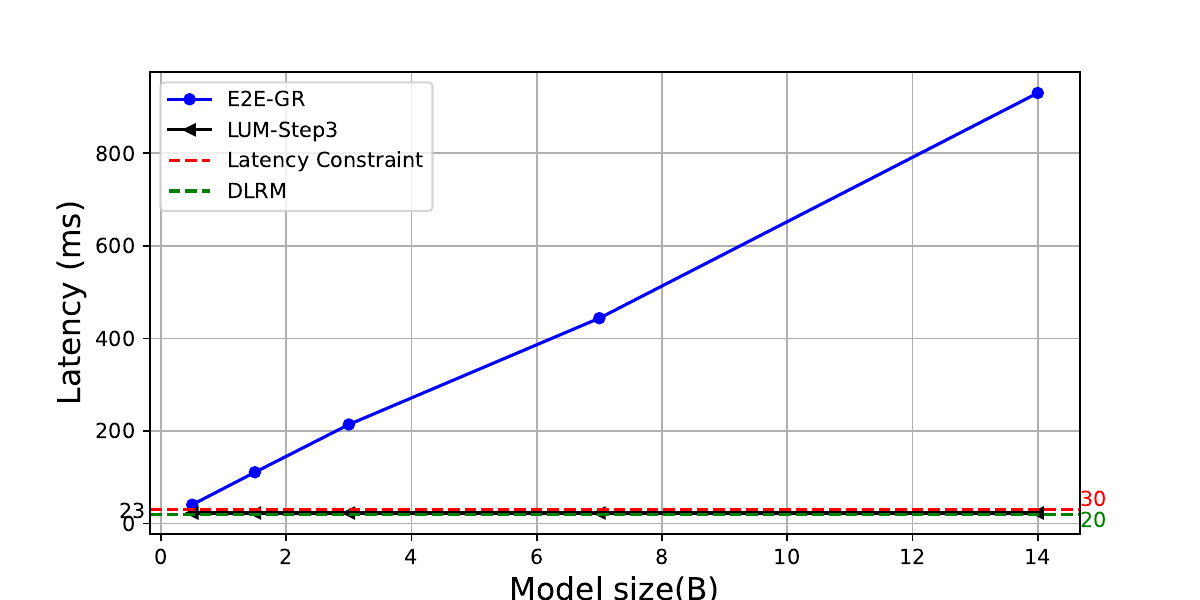}
    \label{fig:subfigure2}
  }
       \subfigure[Max sequence length under latency constrain]{
    \includegraphics[width= .3\textwidth]{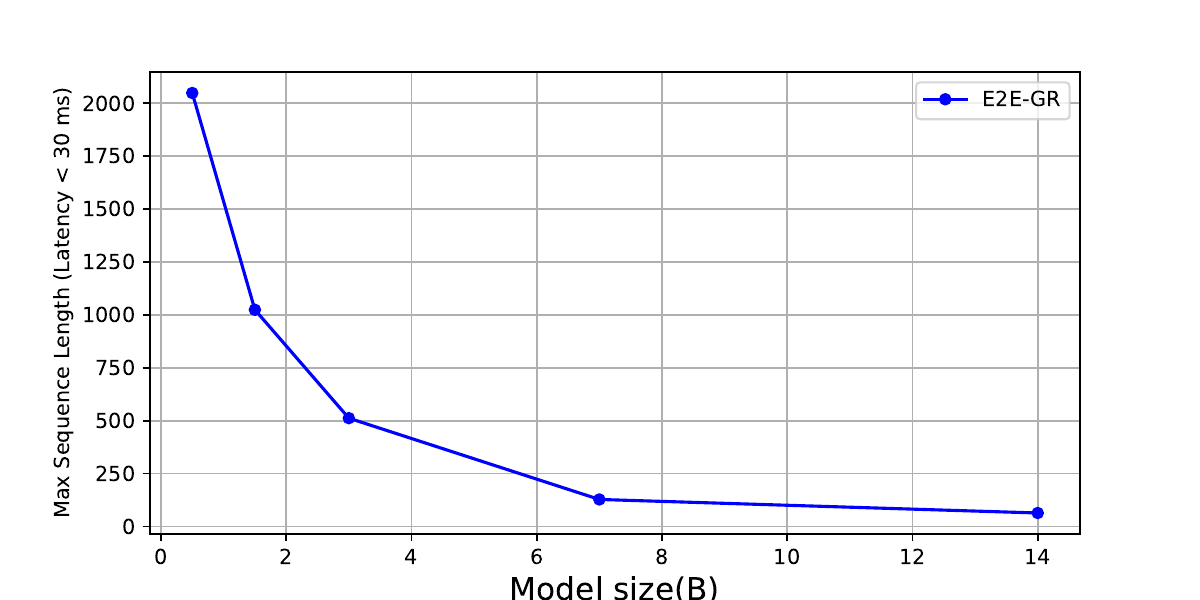}
    \label{fig:subfigure2}
  }
  \vspace{-1em}
\caption{The Results of Efficiency Evaluation.}
\vspace{-1em}
\label{figure:The Results of Efficiency Evaluation.}
\end{figure*}

\subsection{Effectiveness Evaluation}
\label{sec:Effectiveness Evaluation}
In this section, we examine the advantages of the proposed three-step paradigm based LUM.

\noindent \textbf{Impact on various DLRMs.}
Owing to the decomposed design of our proposed paradigm, LUM can be universally integrated into any DLRM during Step 3. 
To evaluate the effectiveness of LUM, we examine its impact on various DLRMs by comparing the performance of the original DLRMs (denoted as Base) and DLRMs augmented with LUM (denoted as Base+LUM) on the ranking task. 
The results are summarized in Table \ref{table:The performance of LUM on various DLRMs.}.
The results demonstrate that, with the assistance of LUM, all methods achieve significant improvements ranging from +0.0053 to +0.0176. 
These findings highlight the versatility and effectiveness of LUM in improving the predictive accuracy of various DLRMs.

\begin{table}[]
\centering
\caption{Results of effectiveness evaluation.
}
\vspace{-1em}
\label{table:The results of effectiveness evaluation.}
\scriptsize
\resizebox{\columnwidth}{!}{
\begin{tabular}{llcc}
\toprule
\textbf{Model}      & \textbf{Training Mode}               &\textbf{AUC}    \\ \midrule
E2E-GR    & training from scratch          &0.7334 \\
DLRM  &training from scratch        & 0.7338 \\
LUM  &training from scratch                & 0.7514 \\ \midrule
DLRM (feature) & incremental training       & 0.7541 \\
DLRM (param)   & incremental training      & 0.7525 \\
DLRM (param+feature) & incremental training& 0.7777 \\ \midrule
LUM (feature)        & incremental training       & 0.7659 \\
LUM (param)           & incremental training      & 0.7620 \\
LUM (param+feature)   & incremental training      & \textbf{0.7794} \\ \midrule
LUM (w/o condition token) &training from scratch  & 0.7416 \\ 
LUM (multi-conditions)   &training from scratch& 0.7545 \\ \midrule
LUM (direct feature)     &training from scratch   & 0.7402 \\  
LUM (direct feature+interest matching) &training from scratch        & 0.7514 \\  \bottomrule
\end{tabular}
}
\end{table}

\noindent \textbf{Compatibility with Online Models}
In industrial applications, online models typically incorporate extensive feature engineering and are continuously updated using billions of interactions per day.  
Existing E2E-GRs often neglect the rich knowledge encoded in these production models—such as learned feature representations or parameter distributions—potentially undermining their performance, especially in mature systems where the online model has been refined over years or even decades.
In contrast, LUM is designed with strong compatibility in mind, enabling seamless integration into incremental training pipelines. Specifically, LUM can inherit either (1) engineered features or (2) model parameters from the continually updated online model—allowing it to leverage historical knowledge and avoid relearning from scratch.
In this section, we evaluate the impact of this compatibility by comparing various configurations of LUM against a representative E2E-GR (HSTU) and the production DLRM (i.e., the online model). We report AUC on an industrial dataset as the primary evaluation metric.


Results are summarized in Table~\ref{table:The results of effectiveness evaluation.} (“scratch”: trained from random initialization; “feature”: with production-grade features; “param”: initialized from the online model). We observe:
(1) LUM variants under incremental training—LUM (feature), (param), and (feature+param) — outperform the scratch baseline by +0.0106 to +0.0280 in AUC, showing strong benefits from knowledge transfer.
(2) While HSTU surpasses DLRM from scratch, it falls well short of the enhanced DLRM (feature+param) (0.7334 vs. 0.7777), underscoring the importance of integration with deployed systems. The limited adaptability of E2E-GRs to production knowledge may hinder real-world deployment.

\noindent \textbf{Impact of the proposed tokenization.}
We evaluate the effectiveness of the proposed tokenization. 
Specifically, we develop a LUM (w/o condition token), and report AUC in Table \ref{table:The results of effectiveness evaluation.}.
The results show that compared with LUM (w/o condition token), LUM achieves better performance due to a better understanding of UBS via given conditions.
Furthermore, we also evaluate the effect of using multiply conditions (Section \ref{sec:Step 2: Knowledge Querying with Given Conditions}) including scenario condition and search term condition, denoted as LUM (multi-conditions).
From Table \ref{table:The results of effectiveness evaluation.}, we can observe adding more conditions can further improve the performance, which shows the potential in terms of performance.

\noindent \textbf{Impact of knowledge utilization.}
We evaluate the different strategies for utilizing knowledge in Step 3. 
The results are presented in Table \ref{table:The results of effectiveness evaluation.}. 
Here, "direct feature" and "interest matching" refer to different strategies of knowledge utilization, as detailed in Section \ref{sec:Step 3: Utilize knowledge in DLRMs}. 
Both LUM (direct feature) and LUM (direct feature + interest matching) achieve significant improvements over DLRM, demonstrating the effectiveness of the proposed strategies.

\begin{table}[] 
\centering
\caption{The impact of group query and packing.
}
\vspace{-1em}
\label{table:The impact of Group query and packing}
\scriptsize
\resizebox{\columnwidth}{!}{
\begin{tabular}{lccc}
\toprule
            & \textbf{Phase}                   & \textbf{Time cost (hour)} & \textbf{Imp.}\\ \midrule

LUM (w/o packing) & \multirow{2}{*}{Step 1} &   151    &   - \\
LUM (w/ packing)   &                         &   26   &   \textbf{+82\%} \\ \midrule
LUM (w/o group query) & \multirow{2}{*}{Step 2} &  17   &  -    \\
LUM (w/ group query)   &                         &   3.6  &  \textbf{+78\%}    \\ \bottomrule
\end{tabular}
}
\end{table}

\noindent \textbf{Impact of packing and group query.}
Packing and group query are designed to accelerate Step 1 and Step 2, respectively.
As shown in Table~\ref{table:The impact of Group query and packing}, these optimizations significantly reduce the computational cost of processing one day's industrial data. Specifically, Step 1 and Step 2 achieve speedups of 82\% and 78\%, respectively, demonstrating their effectiveness in large-scale production environments.

\subsection{Efficiency Evaluation}
\label{sec:Efficiency Evaluation}

\subsubsection{Training Efficiency}
\label{sec:training-efficiency}

\noindent \textbf{Setup.}
For E2E-GR, we follow HSTU~\cite{zhai2024actions}, training from impression-level to user-level. 
The DLRM baseline is our production model and also serves as the backbone for LUM in Step 3. 
Both LUM and E2E-GR use a sequence length of 4096 and scale from 0.5B to 14B parameters, trained on 128 GPUs. 
In industrial streaming training, models must process daily data within 24 hours. For E2E-GR (and DLRM), end-to-end training time defines the cost. For LUM, only Step 3 affects downstream latency due to the decoupled three-step paradigm (Section~\ref{sec:Discussion}), so we report its cost as LUM-Step3. This setup is fair, realistic, and highlights the practical advantages of our framework.

\noindent \textbf{Results.}
Figure~\ref{figure:The Results of Efficiency Evaluation.} (a) compares the training time of different models on one day’s data in a production environment.
LUM-Step3 achieves training efficiency comparable to DLRM and remains largely insensitive to model scale, thanks to its decoupled design.
This property enables scalable training under industrial time constraints, unlocking the potential for scaling laws in real-world systems.
In contrast, E2E-GR is 12$\times$ to 98$\times$ slower than LUM-Step3.
None of the E2E-GR variants (across all sizes) can complete training within the 24-hour deadline.
To match LUM’s throughput, E2E-GR would require 12$\times$ to 98$\times$ more GPUs; to meet the 24-hour lower bound, it would still need 2$\times$ to 18$\times$ the current GPU count.

\subsubsection{Serving Efficiency} 
\textbf{Setup:} 
We follow the M-FALCON setup from HSTU~\cite{zhai2024actions} for E2E-GR.\footnote{Unlike LLMs with static parameters, industrial ranking models are updated in real time, making M-FALCON’s caching strategy inapplicable.} 
The DLRM model serves as both the baseline and LUM backbone. 
We impose a strict inference latency budget of <30 ms per request, with ~100 candidates ranked, reflecting real-world constraints. 
Input sequence length is 4096; model sizes for LUM and E2E-GR range from 0.5B to 14B. As discussed in Section~\ref{sec:Discussion}, serving efficiency hinges on downstream latency. Since Steps 1 and 2 of LUM can be precomputed and cached, we evaluate only Step 3 latency—reported as \textit{LUM-Step3}.

\noindent \textbf{Results.} 
As shown in Figure~\ref{figure:The Results of Efficiency Evaluation.}, LUM achieves constant inference latency across all model sizes, enabling seamless scaling without violating the 30 ms budget.
In contrast, E2E-GR fails to meet the latency requirement even for the smallest model (0.5B parameters).
To explore its feasibility under tighter constraints, we further reduce the sequence length of E2E-GR to satisfy the 30 ms threshold (Figure~\ref{figure:The Results of Efficiency Evaluation.}(c)).
Alarmingly, when using a 14B-parameter model, the maximum sequence length that meets the latency constraint is only 64$\times$ shorter than the standard 4096 used in practice.
These results reveal a fundamental trade-off: although E2E-GR may exhibit favorable scaling behavior in offline evaluations, its real-time applicability in industrial systems is severely limited by inference latency.
In contrast, LUM’s decoupled architecture ensures stable and predictable latency regardless of model size, making it a significantly more practical and scalable solution for real-time industrial recommendation systems.

\begin{figure}[t]
 \centering
     \subfigure[Scaling law with model size]{
    \includegraphics[width= .22\textwidth]{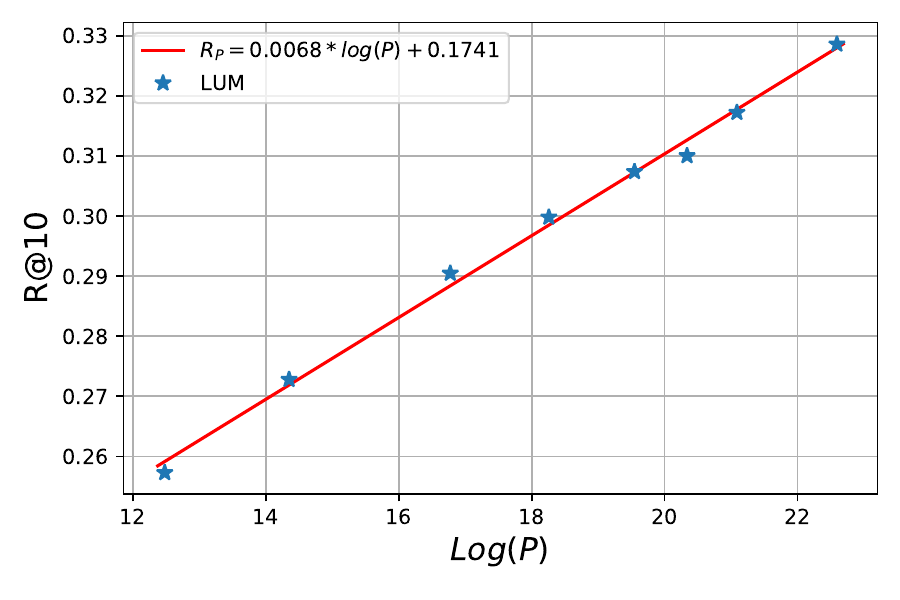}
    \label{fig:subfigure2}
  }
      \subfigure[Scaling law with sequence length]{
    \includegraphics[ width= .22\textwidth]{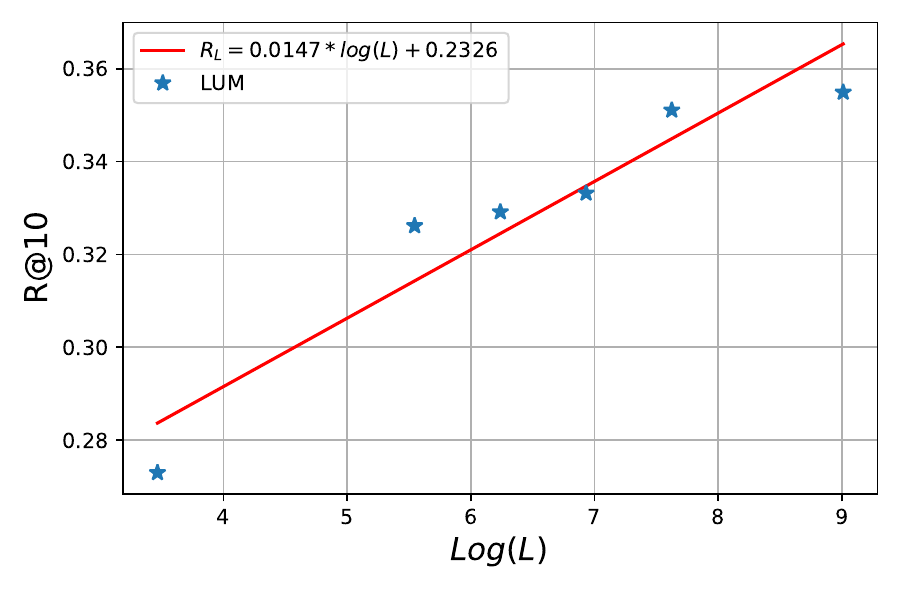}
    \label{fig:subfigure2}
  }
\vspace{-1em}
\caption{Scaling law for LUM.}
\vspace{-1em}
\label{figure:Scaling law for LUM.}
\end{figure}

\vspace{-1em}
\subsection{Scaling Law for LUM}
Following the protocols established in \cite{radford2018improving, radford2019language}, we examine whether LUM complies to similar scaling law. 
For sequence length, to evaluate the impact of sequence length, we train models with varying sequence lengths, from 256 to 8192, under a fixed parameter size of 300 million.
The results are plotted in Figure \ref{figure:Scaling law for LUM.}, where we observe a clear power-law scaling trend, consistent with previous findings \cite{zhai2024actions, radford2018improving, radford2019language}. 
The power-law scaling laws can be expressed as:
\vspace{-0.5em}
\begin{align}
R_P&=0.0068\cdot log(P)+0.1741\\
R_L&=0.0147\cdot log(L)+0.2326 
\vspace{-0.5em}
\end{align} 
where $R_P$ and $R_L$ refer to the R@10 metric for different model sizes and sequence lengths, respectively.
$P$ denotes the model size, and $L$ denotes the sequence length.
These results confirm the strong scalability of LUM, demonstrating that increasing the model size and sequence length can continuously improve the model's performance. 
This finding underscores the potential of LUM to achieve higher performance as it scales, making it a promising approach for large-scale industrial applications.

\subsection{Online Results}
\label{sec:Online Results}
To evaluate the effectiveness of LUM in an industrial setting, we implemented it in the sponsored search system of Taobao, the largest e-commerce platform in China (see Section \ref{sec:Industrial Deployment}). 
We conducted online A/B experiments to test LUM in the ranking task. 
Key performance metrics, CTR and RPM (Revenue Per Mile), demonstrated a significant improvement of 2.9\% and 1.2\% respectively.
These findings highlight the practical benefits of LUM, demonstrating its ability to improve user engagement and business outcomes in large-scale e-commerce platforms. 

\section{Conclusion}
In conclusion, LUM effectively unlocks scaling laws in industrial recommendation systems through a three-step paradigm. 
The decomposed design and next-condition-item prediction ensure that LUM can be efficiently scaled and deployed, leading to significant performance improvements in real-world applications. 
Our experimental and deployment results demonstrate the robustness and practicality of LUM, making it valuable for enhancing user engagement and business outcomes in large-scale e-commerce platforms.

\bibliographystyle{ACM-Reference-Format}


\balance

\end{document}